\documentclass[aps,prb,twocolumn,showpacs,dvipsnames,superscriptaddress]{revtex4-2}

\usepackage[utf8]{inputenc}

\usepackage{amssymb,amsfonts,amsmath} 
\usepackage{graphicx,epsfig,psfrag}
\usepackage{color}
\usepackage{natbib}
\usepackage{url}
\usepackage[breaklinks=true]{hyperref}
\usepackage{mathtools}
\usepackage{subfigure}
\usepackage{physics}
\usepackage{lipsum}
\usepackage{comment}
\hypersetup{
        colorlinks = true,
        citecolor = blue
}
\usepackage{orcidlink}

\begin{document}

\title{Thermal avalanches in a quasiperiodic XXZ model}

\author{Paolo Molignini\orcidlink{0000-0001-6294-3416}}\thanks{paolo.s.molignini@jyu.fi}
\affiliation{Department of Physics and Nanoscience Center, University of Jyväskylä, P.O. Box 35 (YFL), University of Jyväskylä, FI-40014 Jyväskylä, Finland}
\affiliation{Department of Physics, Stockholm University, SE-106 91 Stockholm, Sweden}

\author{Antonio Štrkalj\orcidlink{0000-0002-9062-6001}}\thanks{astrkalj@phy.hr}
\affiliation{\mbox{University of Zagreb Faculty of Science, Department of Physics, Bijenička c. 32, 10000 Zagreb, Croatia}}

\begin{abstract}
We study bath-induced thermalization in the many-body localized XXZ spin chain subjected to a quasiperiodic magnetic field. 
We engineer a thermal inclusion by setting the field strength in one part of the chain below the localization threshold and analyze thermalization via the avalanche mechanism.
To study the nature of such avalanches, we use two complementary observables, namely the two-point connected correlation function and the particle-number entropy.
Surprisingly, the correlations alone show no signatures of avalanches. 
Instead, they display a logarithmic growth of the correlation length throughout the dynamics and predict localization lengths of the local integrals of motion that remain well below the avalanche threshold.
In contrast, the particle number entropy shows clear signatures of the thermal avalanche, with the avalanche front progressively propagating deeper into the localized subsystem for sufficiently large baths.
The discrepancy between the two observables shows that two-point correlations are unreliable to identify thermal avalanches in quasiperiodic systems, as opposed to the random case. 
On one hand, qualitative results are consistent with the analytical predictions of the standard avalanche theory.
On the other hand, significant quantitative deviations persist, which could be due to short-range resonances generated by the quasiperiodic potential. 
Our results suggest that the standard avalanche framework requires revision to account for the short-range correlations of quasiperiodic potentials.
\end{abstract}
\maketitle

\section{Introduction}
Anderson localization in noninteracting disordered systems~\cite{Anderson1958} laid the foundation for many-body localization (MBL), a mechanism through which isolated interacting quantum systems can evade thermalization~\cite{Altshuler1997, Gornyi2005, Basko2006, Oganesyan2007}.
Generic isolated and nonintegrable many-body systems are expected to thermalize following the eigenstate thermalization hypothesis (ETH)~\cite{Deutsch1991,Srednicki1994,Srednicki1999,Rigol2008,DAlessio2016}, according to which individual many-body eigenstates reproduce the thermal expectation values of local observables.
In MBL systems, localization can occur throughout the full many-body spectrum, including at high energy densities formally corresponding to infinite temperature.

A prototypical setting is an interacting spin-$1/2$ chain subjected to a random field.
As the field strength is increased, finite systems can cross over from an ergodic regime satisfying ETH to a localized regime that violates it~\cite{Gopalakrishnan2020}.
This localized regime is phenomenologically described by an extensive set of approximately Local Integrals Of Motion (LIOMs)~\cite{Huse2014, Serbyn2014, Ros2015, Nandkishore2015, Imbrie2016, Thomson2018}, rather than by the global conserved quantities characteristic of conventional integrable systems.
The LIOM framework accounts for several characteristic signatures of MBL, including suppressed transport, saturation of the particle-number entropy, area-law entanglement of many-body eigenstates, and logarithmic growth of the entanglement entropy following a quench~\cite{Znidaric2016, Bardarson2012, Serbyn2013, Iemini2016, Singh2016}.

\begin{figure}[t!]
	\centering
    \includegraphics[width=\columnwidth]{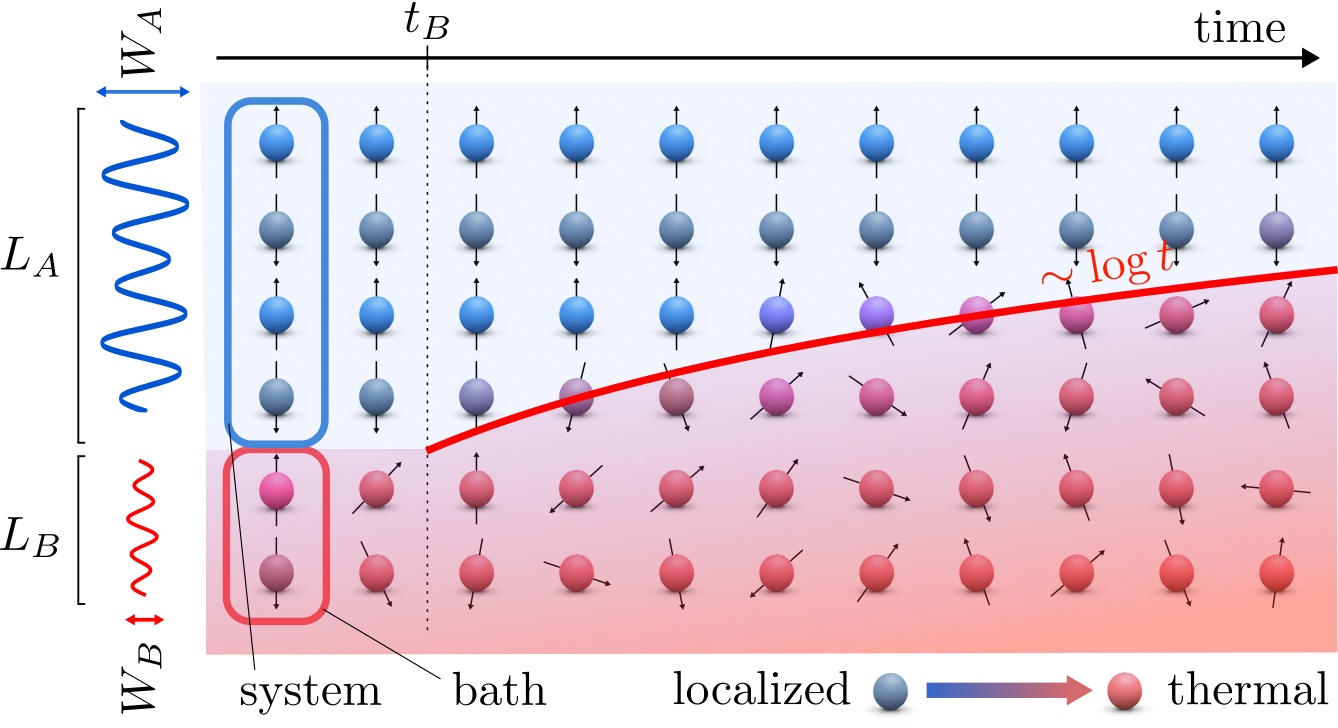}
    \caption{\textbf{Thermal bubble spreading in a quasiperiodic spin chain}. A localized subsystem $A$ (blue, quasiperiodic amplitude $W_A$) is coupled to an ergodic bath $B$ (red, quasiperiodic amplitude $W_B$). The bath acts as a seed for thermalization, generating  after time $t_B$ a thermal bubble that gradually invades subsystem A. The red curve indicates the position of the thermalization front, which propagates approximately logarithmically in time.}
    \label{fig:sketch}
\end{figure}
For finite systems and experimentally accessible timescales, MBL provides a paradigmatic alternative to ergodic dynamics, although its stability in the thermodynamic limit remains under debate~\cite{Nandkishore2015, Abanin2019, Gopalakrishnan2019}.
In particular, the stability of the MBL phase in the thermodynamic limit has been challenged by so-called quantum avalanches that stem from rare \emph{Griffiths regions}~\cite{Griffiths1969, Vojta2010, Agarwal2017, DeRoeck2017b}.
These are atypically large spatial regions in which the disorder is substantially weaker than in the surrounding system.
Although such regions are rare at any finite system size, in an infinite randomly disordered system the probability of finding an arbitrarily large weak-disorder region approaches unity.
Such a region can become ergodic and act as a thermal inclusion that begins to thermalize its localized surroundings.
If the inclusion is sufficiently large that its coupling to neighboring localized degrees of freedom exceeds its many-body level spacing, it can absorb them and grow further.
This self-sustaining process can escalate into an avalanche and ultimately destroy localization throughout the system.

The avalanche mechanism in randomly disordered MBL systems was first formulated analytically~\cite{DeRoeck2017a,Thiery2018} and was subsequently investigated numerically in controlled settings containing finite thermal inclusions
~\cite{Luitz2017, Goihl2019, Morningstar2021, Colmenarez2024, Szoldra2024}.
Avalanche-based ideas have since been incorporated into phenomenological and real-space renormalization-group descriptions of the MBL transition~\cite{Goremykina2019, Dumitrescu2019, Morningstar2019, Crowley2022, Scocco2024, DeRoeck2023}.
Nonetheless, their applicability beyond these simplified settings remains an important open question, particularly for experimentally motivated finite baths~\cite{Sels2022, Ha2023, Leonard2023}, higher-dimensional 
systems~\cite{Potirniche2019, Agrawal2022, Strkalj2022}, and more structured form of disorder~\cite{Doggen2019,Tu2023}.

Most studies of avalanches in MBL systems have been carried out using models with random disorder, where rare weak-disorder regions can arise naturally.
In this context, exact diagonalization (ED) approaches that study the full spectrum are limited to small systems of up to around 20 sites.
At the same time, methods that can access larger one-dimensional systems (such as tensor networks) do so at the expense of reduced numerical accuracy and with system sizes that still remain far from the thermodynamic limit~\cite{Doggen2018, Chanda2020}.
Therefore, avalanches are often studied using implanted thermal inclusions, namely, deliberately engineered weak-disorder regions that act as finite thermal baths~\cite{Luitz2017, Goihl2019, Sels2022, Leonard2023, Colmenarez2024}.

On the other hand, there exists another class of systems that can host MBL, namely, quasiperiodic systems.
One way to induce quasiperiodicity is to modulate the system on-site potential if dealing with particles, or the local magnetic field in case of spins, with a periodic function that is incommensurate with the underlying lattice.  
Such quasiperiodic systems retain the key ingredients needed for localization and MBL, while offering greater flexibility that can bring otherwise higher-dimensional phenomena into simpler settings, as exemplified by single-particle mobility edges in one-dimensional quasiperiodic models compared with their three-dimensional counterparts in the standard Anderson model~\cite{Abrahams1979, Biddle2010, Ganeshan2015, Lueschen2018, An2021, Molignini2025,Goblot2020,Strkalj2021}.
Although quasiperiodic systems, being perfectly deterministic, do not host Griffiths regions, there is growing interest in studying avalanches induced by implanted thermal inclusions in this class of systems~\cite{Tu2023}, especially since quasiperiodic potentials can be readily, controllably, and reproducibly engineered in ultracold atomic platforms~\cite{Leonard2023}.
Clarifying the role of avalanches in quasiperiodic systems is thus essential for understanding the general stability of MBL and for establishing a direct connection between theoretical predictions and experiments.

Motivated by these considerations, in this work we explore the avalanche mechanism in the XXZ chain subjected to an external quasiperiodic magnetic field.
We engineer a thermal inclusion by setting the field strength in one part of the chain within the ergodic regime, such that it acts as a finite thermal bath, see Fig.~\ref{fig:sketch}.
Using two complementary and experimentally relevant observables~\cite{Leonard2023} --- the two-point correlation function and the particle-number entropy --- we study the bath-induced thermal avalanche across a broad range of potential strengths and bath sizes.
Our central finding is that these two observables give strikingly different pictures of the avalanche dynamics: while the correlations suggest that the avalanche terminates throughout the studied parameter regime, the particle-number entropy reveals clear avalanche proliferation below a certain potential strength.
This unexpected discrepancy shows that two-point correlations alone may fail to identify thermal avalanches in quasiperiodic systems and highlights the particle-number entropy as a more sensitive probe of bath-induced transport.
We argue that this behavior originates from the short-range correlations and resonances generated by the quasiperiodic potential, implying that the standard avalanche theory~\cite{DeRoeck2017a} must be refined to properly describe quasiperiodic systems.
Together, these results reveal the limits of correlation-based avalanche diagnostics and motivate extending this analysis to other systems where avalanche mechanism and the LIOM picture may be qualitatively altered, such as long-range interacting systems.

The rest of the paper is organized as follows.
In Sec.~\ref{sec:model}, we introduce the model, while in Sec.~\ref{sec:avalanche} we review the main predictions of avalanche theory.
Section~\ref{sec:methods} presents the numerical methods and observables, followed by the presentation and discussion of our results in Sec.~\ref{sec:results}.
We conclude with a summary and outlook in Sec.~\ref{sec:conclusions}.

\section{Model}
\label{sec:model}
We consider an XXZ spin-$\tfrac12$ chain in a quasiperiodic magnetic field.
The full chain has total length $L$ but it is partitioned into a subsystem $A$ of length $L_A$ and a bath $B$ of length $L_B$, see Fig.~\ref{fig:sketch}, yielding the Hamiltonian 
\begin{equation}    \label{eq:Ham1}
\hat H = \hat H_{\mathrm{int}} + \hat H_{\mathrm{field}}.
\end{equation}

The first term encapsulates the interacting spin chain and reads
\begin{equation}
\hat H_{\mathrm{int}} = J
\sum_{\langle i,j\rangle}
\left[ \hat S_i^x \hat S_j^x + \hat S_i^y \hat S_j^y + \Delta \hat S_i^z \hat S_j^z
\right] \, .
\end{equation}
%
Unless stated otherwise, throughout this work we will set the unit of energy to be $J$ (i.e. setting $J=1$) and fix the interaction strenghts to be uniform across the two regions, i.e. $\Delta=1$.

The magnetic field term is
\begin{equation}
\hat H_{\mathrm{field}} = \sum_{i=1}^L h_i \hat S_i^z.
\end{equation}
For sites $j\in A$, the field is quasiperiodic,
\begin{equation}
h_j = W_A \cos\!\left(2\pi \beta j + \phi_A \right),
\end{equation}
while for $j'\in B$ it is either quasiperiodic,
\begin{equation}
h_{j'} = W_B \cos\!\left(2\pi \beta j' + \phi_B \right),
\end{equation}
or 
uniformly random $h_{j'} \in [-W_B, W_B]$.
In the main text, we will exclusively focus on the case of a quasiperiodic bath, while results for the Anderson bath are shown in Appendix~\ref{app:Anderson}.
In all simulations, we fix the bath disorder strength to $W_B=0.5$, corresponding to a weakly disordered and consequently an ergodic regime.
At such low disorder, the bath rapidly thermalizes and does not exhibit localization effects, thereby providing an efficient thermal reservoir that is not pinned by the potential. 
Note that the Jordan-Wigner transformation~\cite{Jordan1928} maps the Hamiltonian~\eqref{eq:Ham1} to an equivalent fermionic model with conserved particle number; hence, we interchangeably use the terms ``spin'' and ``particle''.

To probe different dynamical regimes of the subsystem $A$, we systematically vary the quasiperiodic field strength $W_A$, which allows us to interpolate between an ergodic regime at small $W_A$, and a strongly localized, MBL-like regime at large $W_A$~\cite{Iyer2013, Doggen2019, Aramthottil2021, Falcao2024}.
The incommensurate frequencies in the quasiperiodic potentials are chosen identical in both subsystems, $\beta=(\sqrt{5}-1)/2$ (the inverse golden ratio).
This choice maximally suppresses commensurability effects and ensures the absence of accidental periodicity, thereby mimicking the role of true disorder while retaining a deterministic potential. 
Finally, the phases $\phi_A$ and $\phi_B$ are sampled uniformly at random in $[0,2\pi)$ for each realization.

\section{Avalanche theory}
\label{sec:avalanche}
Before discussing the observables and the results for the quasiperiodic system, let us review the basics of the analytical predictions for the avalanche theory developed in Ref.~\cite{DeRoeck2017a}. 
For that purpose, it is instructive to rearrange the terms in Hamiltonian~\eqref{eq:Ham1} and divide them into the terms describing the localized part of the system, $H_A$, the bath, $H_B$ and the coupling between the two, $H_{AB}$: 
\begin{equation}
    \hat H = \hat H_A + \hat H_B + \hat H_{AB} \, . 
\end{equation}
The MBL Hamiltonian, $H_A$, can be written in terms of a set of quasilocal integrals of motion (LIOMs) formed by operators $\{\hat \tau_j^z\}$, with $j = 1, ..., L_A$, that are mutually commuting, $[ \hat \tau_j, \hat \tau_k ]=0$. 
The operators $\{\hat \tau_j^z\}$ are obtained by applying a unitary transformation to original spin operators and their expansion can be written as~\cite{Ros2015, Abanin2019}
\begin{equation}
    \hat \tau^z_j = \zeta \hat S^z_j + \sum_{n=1}^{\infty} \Omega_j^{(n)} \hat O_j^{(n)} \, ,
\end{equation}
where $\zeta$ is the finite overlap with the spin at the site $j$ and $O_j^{(n)}$ can contain operators from sites $j-n,...,j,...,j+n$. 
The effect of the operator $\hat \tau^z_j$ on a spin located at a distant site $k$ is exponentially small, which is ensured by $\Omega_j^{(n)} \sim e^{-|j-k|/\xi}$. 
Therefore, the locality of $\hat \tau^z_j$ is controlled by the length scale $\xi$, which can be viewed as the LIOM localization length. 
Note that this analysis holds for an essentially infinite system $A$, while in finite systems, a weak coupling between different LIOMS could be induced by the boundaries.

Following Ref.~\cite{DeRoeck2017a}, we write the whole Hamiltonian as
\begin{equation}    \label{eq:HAB}
    \hat H \approx \hat H_B + 
    \underbrace{\sum_{j=1}^{L_A} \tilde h_j \hat \tau^z_j}_{H_A} + 
    \underbrace{\sum_{j=1}^{L_A} v_{i,j} \hat O_{i} \hat \tau^-_j}_{H_{AB}} \, ,
\end{equation}
where the first term denotes the Hamiltonian of the thermal bath --- in our case given by the $L_B$ spins that feel a weak potential --- and in the second term we kept only the leading, local contribution in the LIOM Hamiltonian. 
The third term is the simplest coupling of the LIOMs in the localized part $A$ to the bath operator $\hat O_{i}$ from part $B$, with an exponentially decaying strength $v_{i,j} = v_0 e^{-\Delta x/\xi}$.
The quantity $\Delta x=|j-i|$ is the distance between the LIOM centered at site $j$ and the bath starting at site $i$. 
Note that all the subleading terms that include higher order products of $\hat \tau^z_j$ operators are neglected in $\hat H_A$ and the LIOM-bath coupling occurs only via the local bath operator $\hat O_i$ located at the site closest to the localized part. 

The exponentially decaying LIOM--bath coupling must be compared with the rapidly increasing density of states of the ergodic region $B$. 
Since the weakly modulated subsystem $B$ acts as a thermal bath, its density of states scales to leading order as $\rho_B \sim 2^{L_B}$.
According to ETH~\cite{Deutsch1991, Srednicki1994, Srednicki1999}, the matrix elements of the bath operator in the eigenbasis $H_B\ket{\psi_n}=E_n\ket{\psi_n}$ are given by
\begin{equation}
    \bra{\psi_n} \hat O_i \ket{\psi_m} = O(\overline{E}) \delta_{n,m}+ \rho_B^{-1/2} f(\overline{E},\omega) R_{n,m} \, , 
\end{equation}
with $\overline{E}=(E_n+E_m)/2$, $\omega=E_n-E_m$ and $R_{n,m}$ being a matrix random variable with zero mean and unit variance. 
We assume that $O(\overline{E})$ and $f(\overline{E},\omega)$ are smooth functions and for simplicity take them to be constants over the probed energy interval. 

Now we are in a position to estimate the thermalization time of the LIOM centered at the site $j$. 
We concentrate on the process that changes $\ket{\Psi_{I}} =  \ket{..., \tau^z_j,...} \otimes \ket{\psi_n}$ to $\Psi_{F} = \ket{..., \overline{\tau^z_j},...} \otimes \ket{\psi_m}$, where $\overline{\tau^z_j}$ denotes the flipped LIOM at site $j$, and $\ket{\psi_{n/m}}$ are the eigenstates of the bath. 
From Fermi's golden rule, it is possible to obtain the rate for such a process 
\begin{equation}
    \Gamma_j \propto \rho_B \, |\bra{\Psi_F} H_{AB} \ket{\Psi_I}|^2 \propto v_{i,j}^2 = v_0^2 e^{-2\Delta x/\xi} \, .
\end{equation}
It follows that the timescale for such thermalization is $t_j \approx 1/\Gamma_j \propto e^{2\Delta x/\xi}$.

Lastly, let us present a brief argument for the avalanche proliferation phenomenon.
The condition for the bath to thermalize the first (nearest) LIOM is that the matrix element of the LIOM-bath coupling has to be much larger than the typical energy gap $\Delta E$ in the spectrum of the bath, namely $G_1 \equiv \bra{\Psi_F} H_{AB} \ket{\Psi_I}/\Delta E \approx v_0 e^{-1/\xi} \rho_B^{1/2} \gg 1$, such that the LIOM sees the bath as a continuous set of energy levels and hybridizes with a large number of bath eigenstates.
Crucially, after the first LIOM has been hybridized, the bath grows and its density of states increases by factor 2, i.e. $\rho_B \sim 2^{L_B} \rightarrow \rho_{B'} \sim 2^{L_B+1}$. 
This enlarged bath has a smaller many-body level spacing and is therefore even more effective at hybridizing the next LIOM, generating the positive feedback that drives the runaway expansion of the thermal region.

After $\Delta x$ LIOMs have been hybridized, $\rho_{B'} \sim 2^{L_B+\Delta x}$, and the hybridization condition becomes
\begin{equation}
    G_{\Delta x} \propto v_0 e^{-\frac{\Delta x}{\xi}} 2^{\frac{L_B+\Delta x}{2}}
    = v_0 2^{\frac{L_B}{2}} e^{\Delta x \left( -\frac{1}{\xi} + \frac{\ln 2}{2} \right)} \, .
\end{equation}
Consequently, the localization is destroyed by the avalanche mechanism when the LIOM localization length is $\xi>\xi_{\rm crit}=2/\ln2\approx 2.9$.

\section{Methods and Observables}
\label{sec:methods}

Having established the analytical expectations for avalanche proliferation, we now introduce the computational protocol used to test them.
We calculate the full time evolution of the spin model and extract observables at each time step. 
Since we consider a quasiperiodic system, there is no intrinsic disorder averaging as in truly random models. 
Instead, we generate an ensemble of realizations by randomizing the phase offsets $\phi_A$ and $\phi_B$ of the quasiperiodic potential. 
This procedure effectively samples different spatial configurations of the incommensurate field while preserving its deterministic structure.

Quasiperiodic systems are known to exhibit partial self-averaging properties, in the sense that fluctuations between different realizations are significantly reduced compared to fully random disorder~\cite{Schreiber2015, Lueschen2018}. 
As a consequence, reliable statistical estimates can be obtained with a relatively moderate number of realizations. 
In practice, we typically average observables over $1024 - 3072$ realizations, which we find sufficient to ensure convergence of all quantities of interest.
In the following, for the ease of notation, we explicitly omit disorder averaging signs.
Every quantity presented hereafter is to be understood as averaged over many different samples.

To determine the dynamics, we employ exact diagonalization up to $L=18$ sites, implemented in the \texttt{EDITH} software~\cite{EDITH}. 
The Hamiltonian is constructed as a sparse matrix in a fixed total magnetization sector $S^z_{\mathrm{tot}} = \sum_{i=1}^L \hat S_i^z=0$.
To generate the time evolution, we first obtain the full spectrum using a standard \texttt{LAPACK} Hermitian eigensolver based on Householder tridiagonalization combined with a divide-and-conquer algorithm.
This procedure yields the full set of eigenenergies $\{E_n\}$ and eigenstates $\{|n\rangle\}$.
The time-evolved state is then obtained exactly for arbitrary times as
\begin{equation}
|\psi(t)\rangle = \sum_n e^{-iE_n t} c_n |n\rangle,
\end{equation}
where $|\psi(0)\rangle = \sum_n c_n |n\rangle$ is the initial state decomposed in the many-body basis and $c_n = \langle n | \psi(0)\rangle$. 
In all our plots, we use a N\'eel state as the initial state. Besides being a simple experimentally accessible product state in the zero-magnetization sector, it provides a pronounced initial spatial modulation for tracking transport and relaxation. 
It also has substantial overlap with many many-body eigenstates, thereby probing the dynamics across a broad energy window.

From the many-body state at time $t$ we can compute arbitrary observables by computing expectation values of different operators.
In this work, we focus on the local $z$-magnetization (spin expectation along the $z$ direction), defined as
\begin{equation}
m_i(t) = \langle \psi(t) | \hat S_i^z | \psi(t) \rangle,
\end{equation}
and on the connected component of the equal-time two-point correlation function,  defined as
\begin{equation}    
\label{eq:conn_correlations}
C_{ij,\mathrm{conn}}^{zz}(t) = C_{ij}^{zz}(t) - m_i(t)m_j(t).
\end{equation}
where
\begin{equation}
C_{ij}^{zz}(t) = \langle \psi(t) | \hat S_i^z \hat S_j^z | \psi(t) \rangle.
\end{equation}

To further characterize the spreading of correlations and the influence of the bath $B$ on the subsystem $A$, we introduce coarse-grained observables that average over the degrees of freedom in the bath. 
In particular, we define a bath-averaged correlation profile
\begin{equation}    
\label{eq:g2}
g^{(2)}(i,t) = \frac{1}{L_B} \sum_{j \in B} C_{ij}^{zz}(t),
\end{equation}
which measures the average correlation between a given site $i$ and all sites belonging to the bath. 
Physically, this quantity provides a direct probe of how strongly a local degree of freedom in the system is influenced by the bath. 
Such bath-averaged correlators are commonly used in the study of MBL and quantum avalanche phenomena~\cite{Szoldra2024, Leonard2023}, where the goal is to monitor how a thermal region in the bath spreads into the localized subsystem.

To extract a characteristic length scale associated with this process, we fit the spatial decay of the bath-averaged correlator to an exponential form,
\begin{equation}    \label{eq:g2_exponential_fit}
|g^{(2)}(i,t)| \sim e^\frac{-|i-L_A|}{\xi_d(t)},
\end{equation}
which defines a time-dependent correlation length $\xi_d(t)$. 

The time dependence of the correlation length $\xi_d(t)$ provides a key diagnostic of the underlying dynamical regime by measuring the extent to which the bath has penetrated into subsystem $A$. 
As highlighted in Sec.~\ref{sec:avalanche}, in a perturbative picture based on Fermi’s golden rule the thermalized region is expected to grow logarithmically in time, $\xi_d(t)\sim \log t$, due to the exponential suppression of couplings away from the bath.
This growth eventually saturates when the discreteness of the bath spectrum is resolved~\cite{DeRoeck2017a, Thiery2018}.
If the growth remains unbounded, $\xi_d(t)\sim \log t$ for arbitrarily long times, or becomes faster than logarithmic, $\xi_d(t)\gg \log t$, this signals the onset of avalanche dynamics, in which the bath effectively grows by hybridizing with nearby degrees of freedom and can eventually thermalize the entire system~\cite{DeRoeck2017a, DeRoeck2014, Szoldra2024}.
By contrast, a saturation of $\xi_d(t)$ indicates that the bath fails to destabilize the localized phase and its influence remains confined.

To complement the correlation-based diagnostics, we also introduce entropy measures that quantify the build-up of quantum correlations between the subsystem and the bath. 
Writing the many-body wavefunction as a tensor product over two partitions $P_1$ and $P_2$ (not necessarily corresponding to subsystem $A$ and $B$)
\begin{equation}
|\psi\rangle = \sum_{p_1, p_2} \psi_{p_1 p_2} |p_1\rangle_{P_1} \otimes |p_2\rangle_{P_2},
\end{equation}
the reduced density matrix of subsystem $P_1$ is obtained as
\begin{equation}
\rho_{P_1} = \mathrm{Tr}_{P_2} \, |\psi\rangle\langle\psi|.
\end{equation}
The von Neumann entanglement entropy, $
S_E = - \mathrm{Tr}(\rho_{P_1} \log \rho_{P_1})$, provides a global measure of quantum entanglement between $P_1$ and $P_2$. 
In ergodic systems, $S_E$ typically grows rapidly and approaches a volume-law value, whereas in MBL systems it exhibits a much slower, often logarithmic growth in time due to dephasing processes~\cite{Bardarson2012, Serbyn2013, Znidaric2008}.

In systems with a conserved particle number (or total magnetization), the reduced density matrix $\rho_{P_1}$ has a block structure labeled by the number of particles $n_{P_1}$ in subsystem $P_1$~\cite{Ghosh2022,Faulend2026},
\begin{equation}
\rho_{P_1} = \bigoplus_{n_{P_1}} p(n_{P_1})\, \rho_{P_1}^{(n_{P_1})},
\end{equation}
where $p(n_{P_1})$ is the probability of finding $n_{P_1}$ particles in subsystem $P_1$, and $\rho_{P_1}^{(n_{P_1})}$ is the normalized reduced density matrix within the sector of fixed particle number $n_{P_1}$, satisfying $\mathrm{Tr}\,\rho_{P_1}^{(n_{P_1})}=1$.
This structure allows us to decompose the entanglement entropy into two physically distinct contributions, $S_E = S_N + S_C$.
In particular, the particle-number entropy,
\begin{equation}    \label{eq:SN}
S_N = -\sum_{n_{P_1}} p(n_{P_1}) \log p(n_{P_1}),
\end{equation}
quantifies fluctuations of conserved quantities between the subsystem and the bath, and is therefore directly associated with particle or magnetization transport across the boundary. 
The configurational entropy $S_C = -\sum_{n_{P_1}} p(n_{P_1}) \mathrm{Tr}\left(\rho_{P_1}^{(n_{P_1})} \log \rho_{P_1}^{(n_{P_1})}\right)$, instead, captures quantum correlations within each particle-number sector and is sensitive to dephasing and entanglement generation even in the absence of transport.
Monitoring $S_N$ and $S_C$ separately can distinguish transport-driven thermalization from entanglement growth generated purely by dephasing. 
In the present work, we focus on $S_N$ only since our primary aim is to track particle transport from the bath into the localized subsystem and thereby diagnose the spatial propagation of the avalanche.

\section{Results}
\label{sec:results}

\subsection{Two-point correlation function}

\begin{figure}[t!]
	\centering
    \includegraphics[scale=1]{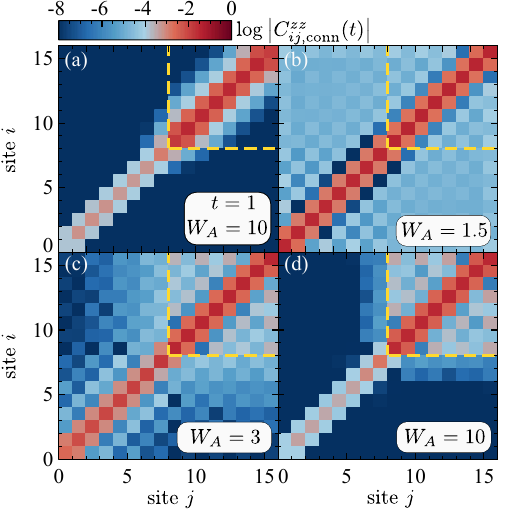}
    \caption{\textbf{Two-point connected correlation function.}
    Logarithm of sample-averaged two-point correlations at short, $t=1$ (a), and long times $t=10000$ (b-d).
    The potential strength in the subsytem $A$ is (a),(d) $W_A=10$, (b) $W=1.5$, and (c) $W=3.0$.
    The times are in units of the hopping $J$.
    In all plots, we used chains with $L=16$ sites and the bath containing $L_B=8$ sites with  weak potential strength $W_B=0.5$. 
    The dynamics is calculated starting from an initial N\'{e}el state. 
    Dashed yellow lines enclose the sites belonging to the bath.
    }
    \label{fig:corr_matrix}
\end{figure}

We start our analysis by studying the time dependence of the two-point connected correlation function $C^{zz}_{ij, {\rm conn}}$, see Eq.~\eqref{eq:conn_correlations}. 
In Fig.~\ref{fig:corr_matrix}, we show the correlation matrix at short times [$t=1$, panel (a)] and long times [$t=10000$, panels (b)-(d)] for different strengths of the potential in the $A$ part, $W_A$, while keeping the bath at $W_B=0.5$. 
For short times, and with a strong potential, the correlations do not extend beyond nearest neighbors in the localized part $A$ and quickly spread only inside the bath $B$, see Fig.~\ref{fig:corr_matrix}(a).
The behavior at long times is shown in Figs.~\ref{fig:corr_matrix}(b-d), and reveals three different dynamical regimes. 

At $W_A=1.5$, the whole system is ergodic, meaning that correlations develop simultaneously across all sites in the chain and become spatially homogeneous.

The opposite behavior is recovered for very high potentials, e.g., $W_A=10$, shown in Fig.~\ref{fig:corr_matrix}(d).
In this case, due to the strong localization of the $A$ subsystem, the correlations remain local even at long timescales: they spread only near the bath and eventually stop. 
Note that there is still a small spreading of correlations between nearest neighbors far away from the bath, comparable to the situation occurring at short times, cf. Fig.~\ref{fig:corr_matrix}(a). 
This residual short-range spreading is consistent with local resonant
hopping processes that are naturally generated by the quasiperiodic
potential~\cite{Faulend2026}.       

Lastly, at $W_A=3$, we observe an intermediate behavior compatible with avalanche dynamics, shown in Fig.~\ref{fig:corr_matrix}(c).
For this potential strength, the isolated subsystem $A$ lies near or inside the finite-size MBL crossover reported for quasiperiodic XXZ chains~\cite{Iyer2013, Doggen2019, Aramthottil2021, Falcao2024}.
In this case, the correlations grow gradually from the bath to more distant sites, eventually thermalizing the full extent of the $A$ subsystem.

\begin{figure}[t!]
	\centering
    \includegraphics[scale=1]{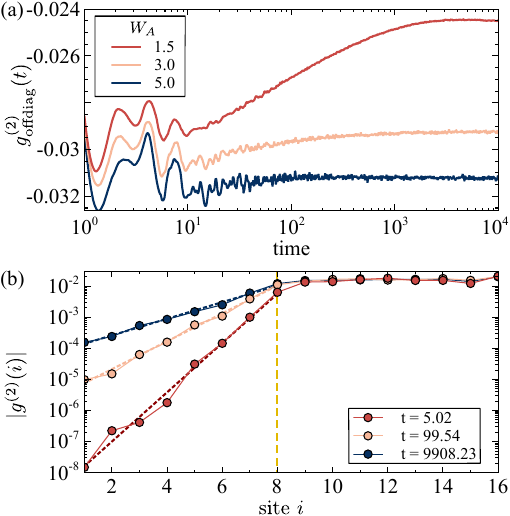}
    \caption{ \textbf{Bath-averaged two-point correlation function and bath thermalization.}
    (a) Mean of the off-diagonal elements of $g^{(2)}$ calculated only in the bath region, i.e. $i,j \in L_B$, for three different strengths of the potential.
    (b) Bath-averaged $g^{(2)}$ as a function of the distance from the bath $i$ plotted for three different hopping times.
    Dashed color-coded lines show the exponential fit inside the localized part of the system. 
    The vertical dashed line shows the position of the last site in the localized subsystem, i.e. $i=L_A$. 
    For all curves we used $L=16, L_A=8,$ and $W_A=3.0, W_B=0.5$.
    All parameters, except $W_A$, are the same as in (a).
    }
    \label{fig:bath_correlations}
\end{figure}

Before analyzing the correlations more quantitatively, we first determine the timescale over which the bath thermalizes, since the bath is also initialized in the N\'{e}el state. 
This establishes whether the bath can already be regarded as a stationary reservoir during the times over which correlations continue to develop in the localized system.
To this end, we calculate the off-diagonal elements of $g^{(2)}$ and plot their average over the bath sites in Fig.~\ref{fig:bath_correlations}(a). 
Aside from initial oscillations that persist up to $t\sim10$, the bath rapidly reaches a stationary (thermal) value for $W_A=3$ and $W_A=5$, with no appreciable evolution of the off-diagonal correlations beyond $\mathcal{O}(10)$ hopping times. 
In contrast, for a potential strength in the ergodic regime, e.g.\ $W_A=1.5$, $g^{(2)}_{\rm offdiag}$ continues to grow up to $\mathcal{O}(1000)$ hopping times. 
We therefore identify the bath thermalization time as $t_B\sim10$ hopping times for values of $W_A$ for which the isolated system is expected to lie in the MBL regime.

With the knowledge of the bath thermalization time, we can now study the time evolution of the bath- and sample-averaged two-point connected correlation function, $|g^{(2)}(i)|$.
Fig.~\ref{fig:bath_correlations}(b) shows its spatial profile for a large bath, $L_B=8$, at three representative times ranging from $t\approx5$ to $t\approx9908$.
At all times, we observe an exponential decay of correlations with the distance from the bath.
Only the slope evolves with time, gradually decreasing as sites in the localized subsystem become more strongly correlated with the bath.
As expected from the separation of timescales, we do not observe a strong back-action of the localized subsystem on the bath.
The exponential spatial profile of $g^{(2)}$ is similar to that previously reported for randomly disordered systems~\cite{Szoldra2024}.
Interestingly, despite the spatial structure of the quasiperiodic potential, we do not resolve corresponding oscillatory features in $g^{(2)}$.
In particular, the bath-averaged correlations do not display the structured signatures associated with local hopping processes in quasiperiodically localized systems~\cite{Faulend2026,Tabanelli2024}.
 
\begin{figure}[t!]
	\centering
    \includegraphics[scale=1]{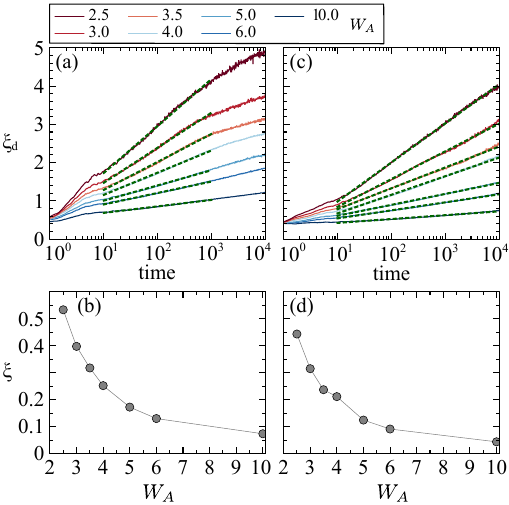}
    \caption{ \textbf{Time dynamics of the correlation length.}
    Correlation length $\xi_d$ as a function of time for different $W_A$ for (a)-(b) a large bath with $L=16, L_B=8$, (c)-(d) a small bath with $L=16, L_B=2$. 
    The dashed lines in (a) and (c) show the exponential fit, from which the LIOM localization length $\xi$ is extracted in (b) and (d).   
    }
    \label{fig:correlation_length}
\end{figure}

From the aforementioned exponential behavior of $|g^{(2)}(i)|$, we can extract at each time the decay length, i.e. the correlation length $\xi_d$ described in Eq.~\eqref{eq:g2_exponential_fit}, using logarithmic linear fitting. 
The results for large and small baths, $L_B=8$ and $L_B=2$ respectively, are shown in Fig.~\ref{fig:correlation_length}(a) and Fig.~\ref{fig:correlation_length}(c) for different potential strengths. 
For the system with a large bath in Fig.~\ref{fig:correlation_length}(a), the short-time dynamics, $t<t_B$ is governed by the thermalization of the bath. 
At $t \sim t_B$, the correlation length $\xi_d$ has a characteristic kink, after which its growth slows down. 
For later times, $t > t_B$, the curves follow a slower, logarithmic growth. 
Note, however, that some curves, i.e. $W_A=\{ 2.5,3.0,3.5\}$ slow down after $\mathcal{O}(1000)$ hopping times.
We attribute this behavior to finite-size effects --- see Appendix~\ref{app:finite-size} for further discussion.
In the case of the smaller bath shown in Fig.~\ref{fig:correlation_length}(c), we observe a similar logarithmic growth, albeit with slower rates than in Fig.~\ref{fig:correlation_length}(a).

We also observe faster-than-logarithmic growth of $\xi_d$ in our numerics.
However, this behavior appears only on the ergodic side of the crossover at lower potential strengths, as shown in Appendix~\ref{app:ergodic}.
This comparison indicates that faster-than-logarithmic growth reflects globally ergodic dynamics in our model, rather than being a necessary signature of avalanche proliferation within the localized regime.

The correlation length $\xi_d$ can be related to the LIOM localization length, as pointed in Ref.~\cite{Szoldra2024}. 
As discussed in Sec.~\ref{sec:avalanche}, the LIOM inside the initially localized part of the system that is $\Delta x$ sites away from the bath, decays after a time $t_{\Delta x} \propto \exp(2\Delta x / \xi)$. 
Taking the distance $\Delta x$ to be given by the correlation length $\xi_d$, we arrive at a simple equation  
\begin{equation}    \label{eq:xi_from_xid}
    \xi_d = \frac{\xi}{2} \, \ln t + {\rm const.} \, .
\end{equation}
Now, it is possible to extract the LIOM localization length $\xi$ by a simple linear fit, shown by the dashed lines in Figs.~\ref{fig:correlation_length}(a) and (c). 
Note that we took special care to exclude the long-time saturation in the case of low $W_A$ and a large bath in Fig.~\ref{fig:correlation_length}(a) by performing a fit only in the interval $t \in [10,1000]$.
The extracted $\xi$ for large and small baths, and as a function of $W_A$, are shown in Figs.~\ref{fig:correlation_length}(b) and (d), respectively. 

As expected, $\xi$ decays with increasing $W_A$ for both cases. 
Changing the bath size mainly renormalizes the slope of the logarithmic growth.
The smaller bath yields slightly lower values of $\xi$, whereas the qualitative time dependence of $\xi_d$ remains unchanged.
Within the correlation-based description, the larger bath therefore has a weak delocalizing effect that increases the effective LIOM localization length, but it does not destroy the LIOM picture.

Surprisingly, all values of $\xi$ lie well below the critical value $\xi_{\rm crit} \approx 2.9$, below which the phenomenological theory of Refs.~\cite{DeRoeck2017a, Thiery2018} predicts that the avalanche mechanism cannot sustain itself, see Sec.~\ref{sec:avalanche}.
Therefore, according to the predictions from the correlations, the avalanche is terminated in the studied system at all $W_A \geq 2.5$, even for a large bath with $L_B=8$ sites, which spans half of the whole system. 

This correlation-based conclusion contrasts with the behavior reported for the randomly disordered XXZ chain in Ref.~\cite{Szoldra2024}, where the correlations show a pronounced evolution from the ergodic regime at $W_A=2.5$, through the critical regime around $W_A=4$, to the localized regime at $W_A=6$.
Within the accessible finite systems, the persistence of the localized correlation profile in our quasiperiodic model is consistent with the sharper and apparently more robust localization crossover reported for quasiperiodic potentials~\cite{Sierant2022, Falcao2024}.

\subsection{Particle-number entropy}
\begin{figure*}[t!]
	\centering
    \includegraphics[scale=1]{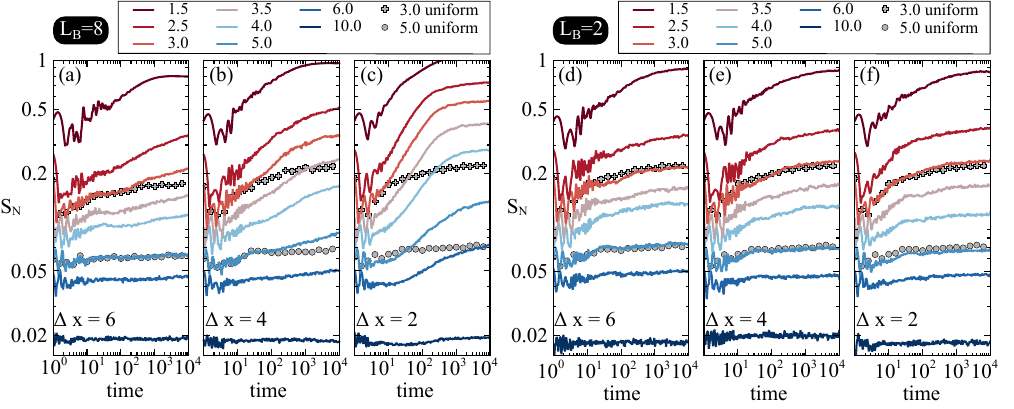}
    \caption{ 
    \textbf{Particle number entropy.}
    Sample-averaged particle number entropy for three different partitions $\Delta x$ and two different bath sizes $L_B=8$ (a-c) and $L_B=2$ (d-f), with different color-coded potential strengths $W_A$.
    The partition boundary is measured from the bath-system interface. 
    Grey lines containing dots and crosses show the uniform system, i.e. system without a bath having $W_B=W_A$, for $W_A=5$ and $W_A=3$, respectively. 
    In (d-f) we use the same data for bath-less system as in (c).  
    The system contains a total of $L=16$ sites.
    }
    \label{fig:particle_entropy}
\end{figure*}

The correlation analysis alleges that the avalanche ultimately terminates, even for the largest bath studied. 
To test this surprising prediction using an independent observable, we now turn to the particle-number entropy $S_N$, defined in Eq.~\eqref{eq:SN}.
Unlike the correlation function, the particle-number entropy directly probes transport through the predefined partition boundary between the two subsystems. 
Therefore, we calculate $S_N$ for three different partitions located at different positions in the localized system $A$, and for different potential strengths $W_A$.
These results are reported in Fig.~\ref{fig:particle_entropy}.

We begin by focusing on a system with a large bath $L_B=8$ and a single partition $\Delta x = 6$ sites from the bath, shown in Fig.~\ref{fig:particle_entropy}(a). 
From the entropy dynamics, it is possible to characterize the $S_N$ curves into three distinct groups.

At weak potential, e.g. $W_A = 1.5$, where the whole system is ergodic, the curves display an immediate growth, which saturates at long hopping times of $\mathcal{O}(1000)$, given by the system size.
This indicates continuous particle transport, as expected for an ergodic system.
Note that the saturation plateau for weaker potential strengths is the finite-size effect.

At the opposite end, for a very strong potential, e.g. $W_A=10$, the curve saturates quickly. This behavior points to the localized nature of the system at the studied timescales. 

The most interesting regime lies in between, at intermediate potential strengths, e.g. $W_A = 3$.
Here, the curve first saturates after $\mathcal{O}(10)$ hopping times, and then resumes to grow at around several hundred hopping times. 
This is a clear signature of the avalanche --- once the avalanche reaches the partition boundary, the particles start to leak into the localized subsystem, increasing the particle number entropy. 

To verify that this secondary growth is induced by the bath rather than by the resonance-induced entropy growth previously reported in localized systems~\cite{Kiefer2020, Ghosh2022, Faulend2026}, we directly compare the dynamics with those of the corresponding system without a bath.
The latter are shown by the gray symbols in Fig.~\ref{fig:particle_entropy}.
The additional growth is observed only in the presence of the bath and can therefore be attributed to the arrival of the avalanche front at the partition boundary.

The propagation of the avalanche front can be inferred by monitoring $S_N$ for partitions located at increasing distances from the bath, shown in Figs.~\ref{fig:correlation_length}(a)-(c). 
This allows us to determine how far bath-induced transport penetrates into the localized subsystem $A$.
For the partition furthest from the bath, $\Delta x=6$, the particle-number entropy exhibits the secondary growth associated with the avalanche for all $W_A\leq4$. 
In contrast, for $W_A=5$, the curves with and without the bath become indistinguishable, indicating that bath-induced transport never reaches this partition because the avalanche dies out before reaching it.
For partitions closer to the bath, the avalanche signature, on the other hand, persists at larger potential strengths. 
In particular, for $\Delta x=2$, the particle-number entropy continues to grow even at $W_A=10$, indicating that the bath is still able to thermalize the sites closest to the interface despite failing to invade the full localized subsystem.

The disagreement with the correlation-based diagnostic is particularly striking at $W_A=3$.
For this potential strength, $S_N$ shows that the avalanche reaches even the most distant partition, whereas the correlation analysis gives $\xi \simeq 0.4$, almost an order of magnitude below the analytical threshold $\xi_{\rm crit} \simeq 2.9$.
Two-point correlations alone are therefore insufficient to determine whether an avalanche proliferates in the quasiperiodic system.
This differs from the correlation-based analysis of the randomly disordered model in Ref.~\cite{Szoldra2024} and shows that, in the present setting, logarithmic growth of $\xi_d$ cannot by itself be interpreted as evidence for the absence of an avalanche, as in the diagnostic used in Ref.~\cite{Leonard2023}.

Reducing the bath size to $L_B=2$ strongly alters the particle-number entropy dynamics, as shown in Fig.~\ref{fig:particle_entropy}. 
The hallmark secondary growth associated with avalanche propagation is now absent for all partition locations. 
Instead, the curves either display the continuous growth expected in the ergodic regime, e.g. for $W_A=1.5$, or closely follow the corresponding curves without the bath, as for $W_A=3$. 
This demonstrates that the bath is too small to sustain an avalanche, and its thermalizing influence remains confined to the sites immediately adjacent to the interface.

\begin{figure}[ht!]
	\centering
    \includegraphics[width=\columnwidth]{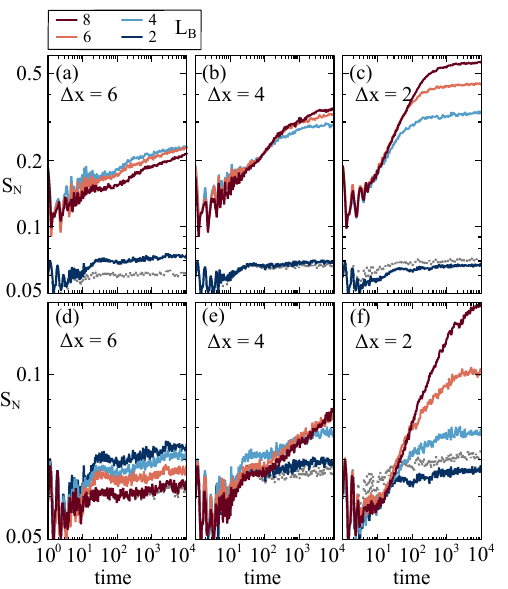}
    \caption{ \textbf{Dependence of the avalanche instability on the bath size.} 
    Sample-averaged particle number entropy (a)-(c) $W_A=3.0$ and (d)-(f) $W_A=5.0$ cases.
    The different panels show results for three different partitions: (a),(d) $\Delta x = 6$, (b),(e) $\Delta x = 4$, and (c),(f) $\Delta x = 2$.
    All systems have $L=16$ sites.
    }
    \label{fig:bath_size}
\end{figure}

In Fig.~\ref{fig:bath_size}, we further investigate the avalanche dependence on the bath size $L_B$ at two different potential strengths, $W_A = \{3.0,5.0\}$, and at the same three partitions $\Delta x \in \{2, 4, 6\}$. 
For the weaker potential, c.f. Fig.~\ref{fig:bath_size}(a)-(c), all bath sizes, except the smallest $L_B=2$, show particle transport across all three partitions: the curves are clearly distinguishable from the $W_B=W_A$ case (gray dotted line) both in magnitude and growth rate. 

On the other hand, for a stronger potential with $W_A=5$, c.f. Fig.~\ref{fig:bath_size}(d)-(f), no significant transport occurs at the furthest partition $\Delta x = 6$ for all studied bath sizes. 
For the partitions closer to the bath, $\Delta x=\{2,4\}$, we observe bath spreading for $L_B=\{6,8\}$ spins but no spreading for $L_B=2$. 
The growth for $L_B=4$ is less obvious, especially at $\Delta x=4$.
Here, the curve has a larger magnitude than in the $W_B=W_A$ case, but the saturation occurs relatively quickly.
Overall, we find that baths with $L_B\ge4$ sustain an avalanche at $W_A=3.0$, whereas at $W_A=5.0$ the avalanche always dies out before reaching the furthest partition.
This qualitative dependence on both the initial bath size and the potential strength is consistent with avalanche theory~\cite{DeRoeck2017a, Thiery2018, Luitz2017}: increasing the initial bath size favors avalanche proliferation, whereas stronger localization restricts the distance over which the thermal inclusion can grow.

\begin{figure}[t!]
	\centering
    \includegraphics[scale=1]{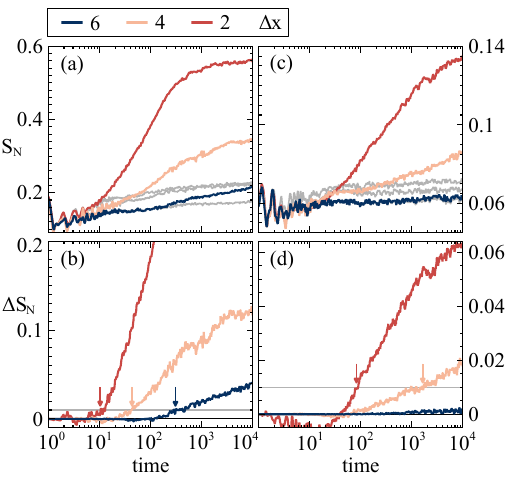}
    \caption{ \textbf{Propagation of the avalanche front.}
    The sample averaged particle number entropy at (a),(b) weak ($W_A=3.0$), and (c),(d) strong ($W_A=5.0$) potentials in the presence of a bath (colored lines) compared to systems without the bath (gray lines). 
    Panels (a),(c) show the full particle number entropy, while (b),(d) show the difference between systems with and without the bath, $\Delta S_N$. 
    Gray horizontal lines in (b) and (d) denote the arbitrary cutoff given by the $\Delta S_N=0.01$. 
    Arrows indicate the time when the avalanche front reached each partition, and are given by the intersection of $\Delta S_N$ curves with horizontal gray lines.
    In all plots, we used systems with $L=16$ sites, out of which $L_B=8$ are bath sites.
    }
    \label{fig:avalanche_front}
\end{figure}

Comparing the particle entropy curves with the uniform system, i.e. without the bath ($W_B=W_A$), also allows us to visualize the propagation of the avalanche front directly.
This is shown in Fig.~\ref{fig:avalanche_front} for the $L_B=8$ site bath and two different potential strengths, $W_A=\{3.0,5.0\}$. 
In Figs.~\ref{fig:avalanche_front}(a) and (c), we explicitly plot both curves (with and without bath) for each case, while in Figs.~\ref{fig:avalanche_front}(b) and (d) we show the difference $\Delta S_N \equiv S_N^{\rm bath} - S_N^{\rm no\,bath}$. 
To estimate the propagation time of the avalanche front, we define an arbitrary threshold value of $\Delta S_N$ equal to $5\%$ of the maximum value of $S_N$ in Figs.~\ref{fig:avalanche_front}(a) and~(c). 
For each partition, we then identify the front-arrival time as the first time at which $\Delta S_N$ crosses this threshold. 
Using this operational definition, we find that the spreading time increases approximately exponentially with the distance from the bath. 
For the furthest partition, $\Delta x=6$, $\Delta S_N$ remains close to zero over the full time window, indicating that the avalanche does not reach this partition, in agreement with the previous results.

Lastly, let us try to crudely estimate the LIOM decay length from the propagation of the avalanche front by using the same exponential dependence of the LIOM decay time, $t_{\Delta x} \propto e^{2\Delta x/\xi_S}$. 
By looking at the approximate positions of the arrows in Figs.~\ref{fig:avalanche_front}(b) and (d), for $W_A=3.0$ we obtain $t_{x=\{2,4,6\}} \sim \{10,50,300\}$, while for $W_A=5.0$ we obtain $t_{x=\{2,4,6\}} \sim \{90,2000,\infty\}$. 
Taking the ratios $t_2/t_4$ and $t_4/t_6$, we estimate $\xi_S$ to be $\xi_S = \{-4/\ln(t_2/t_4),-4/\ln(t_4/t_6)\} = \{2.49,2.23\}$ for the weak potential, and $\xi_S = \{1.29,0\}$ for the strong potential. 
Note that these values are much larger than those obtained from $\xi_d$. 
Furthermore, the first set of $\xi_S$ values is comparable to the critical $\xi_{\rm aval}$ predicted by the analytical arguments, while the latter yields a much smaller value for the first ratio and zero for the second. 
This is in agreement with the fact that the system with $W_A=3.0$ and $L_B=8$ truly hosts an avalanche, while for $W_A=5.0$ the avalanche stops at a certain point.

While the particle-number entropy reproduces the qualitative dependence on bath size and localization strength expected from avalanche theory, we still observe quantitative deviations from its standard predictions.
One possible origin of these deviations would be that the quasiperiodic region with $W_B=0.5$ does not act as the featureless thermal bath assumed in conventional avalanche treatments~\cite{DeRoeck2017a,Luitz2017}.
However, replacing the quasiperiodic potential in the bath with an independently sampled random potential produces no appreciable change in the correlation dynamics, as shown in Appendix~\ref{app:Anderson}.
This suggests that the discrepancy originates primarily from the properties of the localized quasiperiodic subsystem $A$, rather than from those of the bath.

A possible microscopic mechanism is provided by the structured resonances generated by the quasiperiodic potential~\cite{Tabanelli2024, Faulend2026, Padhan2026}.
Such resonances may renormalize the effective LIOM--bath coupling $v_{i,j}$ in Eq.~\eqref{eq:HAB}, causing it to decay more slowly or less uniformly than the simple exponential form assumed in the analytical argument.
This could lower the effective critical localization length for avalanche proliferation and explain why $S_N$ detects an avalanche even when the correlation-based estimates of $\xi$ remain below $\xi_{\rm crit}$.

\section{Conclusions and Outlook}
\label{sec:conclusions}
In this work, we studied the bath-induced thermalization of the quasiperiodic XXZ model.
Unlike in random systems, the quasiperiodic models does not host rare regions of weak potential that would act as a bath, so in order to study the thermal avalanches, we engineered a thermal inclusion by reducing the potential strength below the critical value in one part of the system. 
We performed a detailed analysis of the two-point connected correlation function $g^{(2)}$ and the particle number entropy $S_N$, across a broad range of potential strengths $W_A$ and bath sizes $L_B$. 

From the analysis of correlations, we reveal an exponential decay of $g^{(2)}$, with a decaying length $\xi_d$ growing logarithmically in time for all $W_A$ where the system without the bath ought to be localized. 
The extracted LIOM localization length $\xi$ remains well below the $\xi_{\rm crit}$ predicted by the avalanche theory across the entire parameter range  indicating no presence of the avalanche. 

The particle number entropy $S_N$, however, tells a different story.
For $W_A\leq4$, we observe clear signatures of avalanches, although the correlation analysis indicates otherwise. 
This demonstrates that the two-point correlations alone are insufficient to properly capture the avalanche mechanism in the quasiperiodic systems, as opposed to what reported previously in random system~\cite{Szoldra2024}. 
The qualitative behavior of the avalanche proliferation is observed through $S_N$ and it is consistent with avalanche theory, yet we observe large quantitative deviations from analytical predictions. 
We attribute them to short range resonances in the potential that might lower the predicted $\xi_{\rm crit}$. 

These findings call for a refinement of the avalanche framework of Refs.~\cite{DeRoeck2017a, Thiery2018} that explicitly accounts for the spatial correlations and structured resonances of quasiperiodic systems.
A natural next step is therefore to characterize these resonances more directly and determine how they renormalize the effective coupling between the thermal inclusion and the localized degrees of freedom.

From an experimental perspective, our results show that measurements of correlation spreading should be complemented by the particle-number entropy, or by related transport-sensitive observables, because logarithmic correlation growth alone may incorrectly suggest that an avalanche has terminated.
Such comparisons are already accessible in existing one-dimensional experimental platforms~\cite{Leonard2023}.
It would also be valuable to extend this analysis to higher dimensions, where numerical simulations are considerably more demanding and where the stability of both random and quasiperiodic MBL remains especially uncertain~\cite{Choi2016, Bordia2017, Potirniche2019, Agrawal2022, Strkalj2022,Hur2025}.

Another timely extension would be to investigate how long-range interactions modify the stability of MBL and the avalanche mechanism.
Power-law interactions have already shown to dramatically alter spectral properties of quasiperiodic systems, such as renormalizing single-particle mobility edges~\cite{Molignini2025}.
This can qualitatively reorganize resonant processes and correlation spreading~\cite{Yao2014, Nandkishore2017, Nag2019, Deng2020}, and recent work has begun to address their effect directly on bath-induced avalanche dynamics~\cite{Shen2026}.
A key question to address is whether the conventional LIOM description survives with more extended, possibly algebraically decaying conserved operators, or whether sufficiently long-ranged interactions require a fundamentally different description of the localized regime. 

Since quasiperiodic potentials can be easily generated in ultracold atomic experiments by superimposing optical lattices with incommensurate wavelengths~\cite{Schreiber2015, Lueschen2018, Leonard2023}, these platforms are some of the most natural arenas where to consider the effect of long-range couplings.
In this context, magnetic atoms and polar molecules provide an increasingly versatile toolbox in which dipole--dipole interactions coexist with tunable short-range interactions~\cite{Chomaz2023, Bohn2017}.
Another way to engineer long-range, even effectively infinite-range interactions is via light-matter coupling interactions.
These have already been shown to support long-lived nonergodic regimes in cavity-coupled systems~\cite{Sierant2019, Kubala2021, Chanda2022, Molignini2018, Molignini2022}.

Finally, beyond effective lattice Hamiltonians, these questions could be numerically investigated in the full continuum geometries employed in ultracold-atom experiments using multiconfigurational methods such as MCTDH-X~\cite{Lode2020, Lin2020, Molignini2025MCTDH}.
This would permit a direct treatment of optical potentials and nonlocal interactions, and would build naturally on recent continuum studies of dipolar bosons and fermions in optical lattices~\cite{Molignini2025Bosons, Molignini2025, Molignini2025Quench}.

\begin{acknowledgments}
We thank A. Fritz for useful discussions.
P.M. acknowledges support by the Swedish Research Council (2024-05213).
The work of A.\v{S}. is supported by the European Union’s Horizon Europe research and innovation programme under the Marie Sk\l{}odowska-Curie Actions Grant agreement No. 101104378.
A.\v{S} also acknowledges support from the project “Implementation of cutting-edge research and its application as part of the Scientific Center of Excellence for Quantum and Complex Systems, and Representations of Lie Algebras”, Grant No. PK.1.1.10.0004, co-financed by the European Union through the European Regional Development Fund—Competitiveness and Cohesion Programme 2021-2027, as well as the European Union – NextGenerationEU through the National Recovery and Resilience Plan 2021-2026. Institutional grant of University of Zagreb Faculty of Science ‘’Encouraging competitive projects and top-tier scientific publications at the Department of Physics (ProPuBFO-1.1.3.2026)’’.
Computation time on the Sunrise Compute Cluster of Stockholm University and on the Euler cluster at the High-Performance Computing Center of ETH Zurich is gratefully acknowledged.
The raw simulation data from exact diagonalization is available upon a reasonable request to the authors.
\end{acknowledgments}

\clearpage
\appendix

\section{Finite size effects}
\label{app:finite-size}
%
\begin{figure}[ht!]
	\centering
    \includegraphics[scale=1]{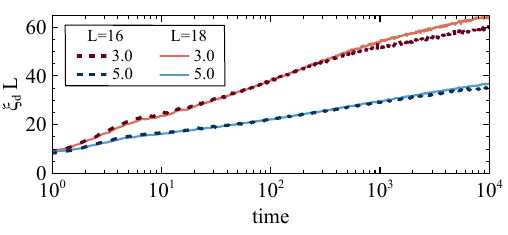}
    \caption{ \textbf{Finite size effect in the correlation length.}
    The time evolution of the sample-averaged correlation length $\xi_d$ scaled by the system size for $W_A=3.0$ (red) and $W_A=5.0$ (blue). 
    We compare a system with $L=16$ sites (darker shades) with a larger system containing $L=18$ sites (lighter shades). 
    All systems have $L_B=8$ sites in the bath. 
    }
    \label{fig:finite_size}
\end{figure}
In this Appendix, we briefly present how the finite size of the system influences the correlation depth $\xi_d$.
In Fig.~\ref{fig:finite_size}, we consider the case of a large bath with $L_B=8$ sites and compare the time evolution of the scaled correlation depth $\xi_d L$ for two system sizes -- $L=16$ and $L=18$ (equivalently $L_A=8$ and $L_A=10$). 
We consider two representative potential strengths, $W_A=3.0$ and $W_A=5.0$.

For $W_A=3.0$, which lies in the regime where the particle-number entropy signals an avalanche reaching the most distant partition [see Fig.~\ref{fig:particle_entropy}(a)], the curves for the two system sizes overlap almost perfectly up to $\sim1000$ hopping times. 
At later times, though, the curve for the larger system grows faster, remaining closer to the $\log t$ growth.
For $W_A=5.0$, where the particle-number entropy indicates instead that the avalanche terminates before reaching the most distant partition, the curves for $L=16$ and $L=18$ remain nearly indistinguishable over the entire time interval.

This comparison indicates that the deviation from $\log t$ behavior at later times, observed for $W_A \leq 4.0$ in Fig.~\ref{fig:correlation_length}(a), results from the thermalization front approaching the finite boundary of the localized subsystem, rather than from a genuine termination of the logarithmic growth.
Consequently, for the system with the large bath, we extract the LIOM localization length $\xi$ by fitting the logarithmic growth only over the interval $t\in[10,1000]$, before finite size corrections begin to separate the curves.

\section{Ergodic phase}
\label{app:ergodic}
%
\begin{figure}[ht!]
	\centering
    \includegraphics[scale=1]{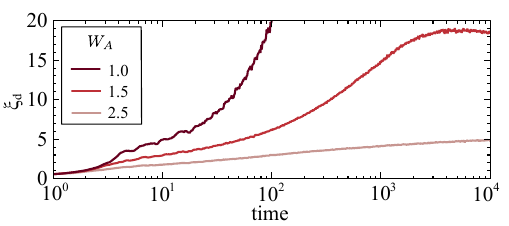}
    \caption{ \textbf{Correlation dynamics in the ergodic side of the transition.}
    The sample averaged correlation length $\xi_d$ for some lower values of the potential $W_A \lesssim 2.5$, showing strong deviations from logarithmic growth. The shared parameters are identical to the ones we used in Fig.~\ref{fig:correlation_length}(a).
    }
    \label{fig:ergodic_correlations}
\end{figure}
In this Appendix, we examine the behavior of the correlation length $\xi_d$ on the ergodic side of the transition, i.e. for $W_A \lesssim 2.5$.
In our analysis of $\xi_d$ in the main text, we focused on the range of $W_A$ for which the system without the bath lies in the MBL regime.
In this regime, we observe logarithmic growth of $\xi_d$ in time throughout the entire parameter range considered.
Taken together with the clear avalanche signatures observed in $S_N$, this leads to a different conclusion from Ref.~\cite{Leonard2023}, where the avalanche is identified through faster-than-logarithmic growth of $\xi_d$.
To observe growth of $\xi_d$ that is faster than $\log t$, we had to reduce the potential strength $W_A$ to values for which the entire system is ergodic, as shown in Fig.~\ref{fig:ergodic_correlations}.
This comparison therefore indicates that faster-than-logarithmic growth of $\xi_d$ reflects globally ergodic dynamics rather than being a necessary signature of avalanche proliferation within the localized regime.

\section{Results for randomly-disordered bath}
\label{app:Anderson}
%
\begin{figure}[ht!]
	\centering
    \includegraphics[scale=1]{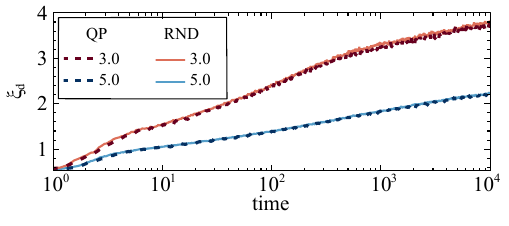}
    \caption{ \textbf{Comparison between quasiperiodic and random bath.}
    The sample averaged correlation length $\xi_d$ for the case of quasiperiodic (QP) and random (RND) baths of size $L_B=8$. 
    The potential strength inside the bath is fixed at $W_B=0.5$, while inside the localized subsystem is either $W_A=3.0$ or $W_A=5.0$, as indicated in the legend.
    Note that the light-blue curve is difficult to distinguish because it lies almost exactly beneath the dark-blue curve.
    }
    \label{fig:random_bath}
\end{figure}
In this Appendix, we verify that our results do not depend significantly on whether the bath is subjected to a quasiperiodic or a random potential.
Since we use a quasiperiodic potential in the bath region, albeit with a weak potential strength $W_B=0.5$, there is a possibility that correlations within the bath could alter its properties, thereby violating the assumption of a featureless bath used in standard avalanche theory~\cite{DeRoeck2017a}.
For this reason, we also calculate the sample-averaged correlation length $\xi_d$ using a random external potential in the bath region, where each field value $h_j \in [-W_B,W_B]$ is independently drawn from a uniform distribution with zero mean.
In Fig.~\ref{fig:random_bath}, we compare the results obtained with quasiperiodic and random baths for two different potential strengths, $W_A=\{3.0,5.0\}$.
We observe no appreciable differences between the two types of bath.

\end{document}